\documentclass[aps,prl,preprint,superscriptaddress]{revtex4}

\usepackage{graphicx}% include figure files
\usepackage{bm}% bold math

\begin{document}

%Title of paper
\title{3D-Chiral Metamaterial Showing Artificial Magnetic Response and Negative Refraction\footnote{As presented at the Quantum Electronics and Laser Sciences Conference (CLEO/QELS 2008), San Jose, CA, USA, 5 May 2008, paper QMA4.}}

\author{E. Plum} \email[Email: ]{erp@orc.soton.ac.uk} \homepage[Homepage:
]{www.nanophotonics.org.uk/niz/} \affiliation{Optoelectronics
Research Centre, University of Southampton, SO17 1BJ, UK}

\author{J. Dong}
\affiliation{Institute of Optical Fiber Commun. and Network Tech.,
Ningbo University, Ningbo 315211, China} \affiliation{Ames
Laboratory and Department of Physics and Astronomy, Iowa State
University, Ames, Iowa 50011, USA}

\author{J. Zhou}
\affiliation{Department of Electrical and Computer Engineering and
Microelectronics Research Center, Iowa State University, Ames, Iowa
50011, USA} \affiliation{Ames Laboratory and Department of Physics
and Astronomy, Iowa State University, Ames, Iowa 50011, USA}

\author{V. A. Fedotov}
\affiliation{Optoelectronics Research Centre, University of
Southampton, SO17 1BJ, UK}

\author{T. Koschny}
\affiliation{Ames Laboratory and Department of Physics and
Astronomy, Iowa State University, Ames, Iowa 50011, USA}
\affiliation{Institute of Electronic Structure and Laser -
Foundation for Research and Technology Hellas (FORTH), and
Department of Materials Science and Technology, University of Crete,
Greece}

\author{C. M. Soukoulis}
\affiliation{Ames Laboratory and Department of Physics and
Astronomy, Iowa State University, Ames, Iowa 50011, USA}
\affiliation{Institute of Electronic Structure and Laser -
Foundation for Research and Technology Hellas (FORTH), and
Department of Materials Science and Technology, University of Crete,
Greece}

\author{N. I. Zheludev}
%\email[Email: ]{n.i.zheludev@soton.ac.uk}
\affiliation{Optoelectronics Research Centre, University of
Southampton, SO17 1BJ, UK}

%\date{\today}

\begin{abstract}
Artificial magnetism, negative permeability and negative refractive
index are demonstrated in 3D-chiral metamaterial that shows giant
polarization rotation and circular dichroism.
\end{abstract}

\maketitle

A simultaneous presence of electric and magnetic responses at
optical frequencies is a necessary condition for any medium to show
optical activity, i.e. differential circular dichroism and
birefringence. However, in natural media such as optically active
crystals and organic liquids the magnetic response is much weaker
than the electric response. Here we demonstrate that artificially
created new optically active 3D metamaterial shows strong resonant
magnetic permeability. The structure exhibits bands of negative
permeability for one and both circular polarizations and a negative
refractive index for one circular polarization. The metamaterial
also shows exceptionally strong circular dichroism and polarization
rotation in relatively wide frequency bands with low transmission
losses, which makes it a very efficient material for circular
polarizers and polarization rotators.

The metamaterial is based on a double-periodic array of layered
3D-chiral meta-molecules (unit cells) formed by pairs of mutually
twisted rosette-like metallic particles, with no contact between
individual layers. Here chiral interaction between the layers is
provided by electromagnetic coupling
\cite{PRL_Rogacheva_2006_GiantGyrotropy, APL_Plum_2007_Gyrotropy}.
Recently we showed experimentally that this type of structures
manifests a ``left-handed" behavior: the phase and group velocities
for one of the circularly polarized eigenstates have opposite signs
indicating the appearance of a backward wave. The latter was noted
as a signature of negative refraction in resonant chiral media, as
predicted by J. Pendry \cite{Science_Pendry_2004_ChiralNegRef}.

\begin{figure}[h!]
\includegraphics[width=160mm]{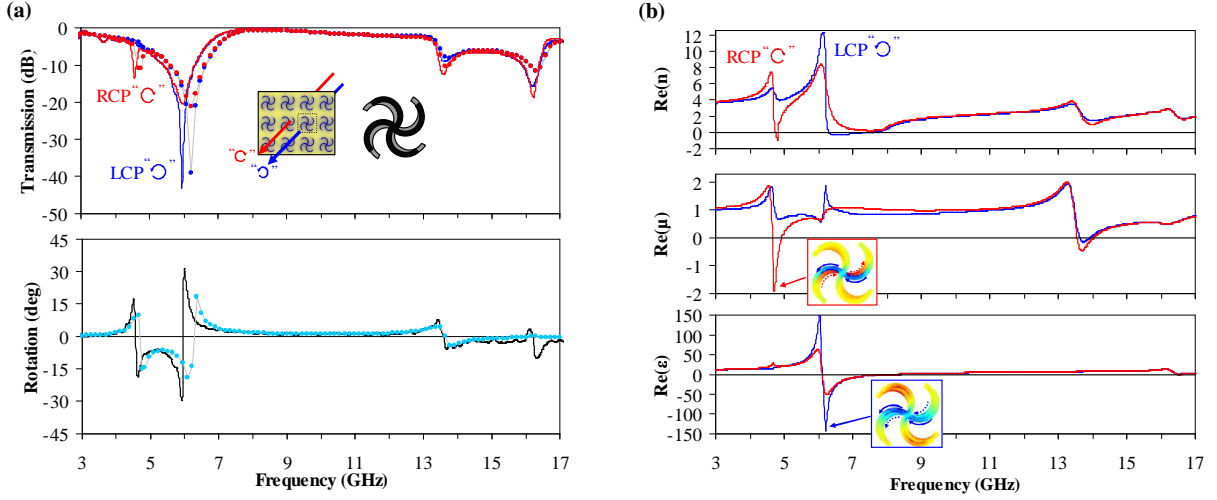}
\caption{\label{fig1}Properties of the bi-layered form of the
metamaterial, which is shown in the left inset. (a) Transmission
levels for right-handed (RCP) and left-handed (LCP) circular
polarizations and azimuth rotation for linear polarization.
Measurements (solid lines) and simulations (points) are shown. (b)
Effective parameters derived from the material's simulated
transmission and reflection properties. The insets show the
horizontal component of the current modes corresponding to the
negative permeability resonance (RCP, middle) and the negative
permittivity resonance (LCP, bottom).}
\end{figure}

\begin{figure}[h!]
\includegraphics[width=160mm]{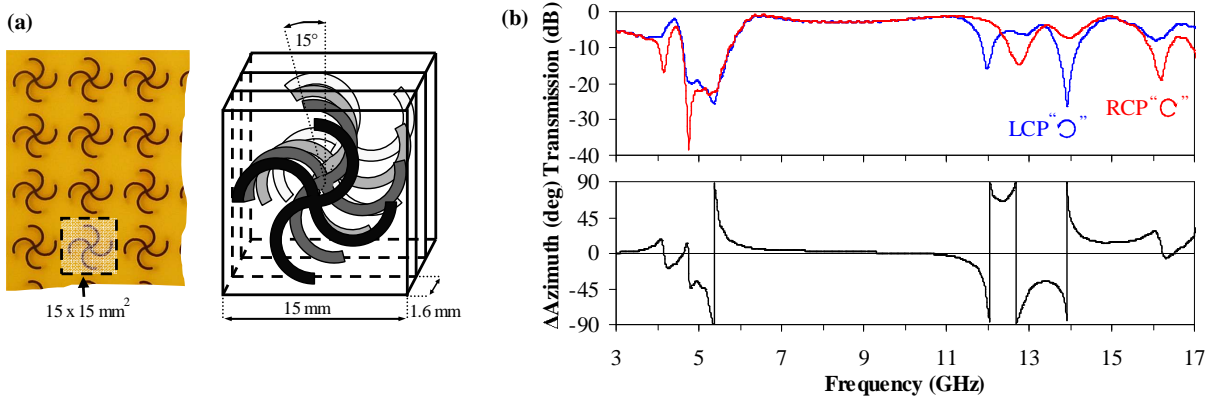}
\caption{\label{fig2}The 4-layered version of the metamaterial. (a)
A fragment of the 4-layered material and schematics of its
three-dimensionally chiral unit cell. (b) Transmission levels for
right-handed (RCP) and left-handed (LCP) circular polarizations and
azimuth rotation for linear polarization.}
\end{figure}

Here we provide evidence that indeed negative refraction takes place
in these structures. We also extend this study to multi-layered
chiral structures. We study three structures with different unit
cells containing correspondingly 2, 4 and 6 coaxial planar copper
rosettes of 4-fold symmetry. The rosettes are located in parallel
planes and separated by very thin (1.6~mm) dielectric layers, which
results in the metamaterial's overall thickness of only 1.6, 4.8 and
8.0~mm correspondingly. A mutual anti-clockwise twist of $15^\circ$
is introduced between adjacent rosettes (see Fig. \ref{fig1}a inset,
Fig. \ref{fig2}a), whose lateral dimension of $15 \times
15~\text{mm}^2$ ensures no diffraction by the metamaterial at
frequencies below 20~GHz.

Fig. \ref{fig1}a shows transmission properties of the bi-layered
form of the metamaterial. For left-handed (LCP) and right-handed
(RCP) circular polarizations the material shows exceptionally strong
circular dichroism of up to 20~dB. For linear polarization azimuth
rotation of up to 30° is achieved. These values are substantial
considering the material's thickness of only 1/30 wavelength
($\lambda$) at 6~GHz where the strongest effects occur. Apart from a
minor frequency shift, numerical results (dots, Fig. \ref{fig1}a)
are in excellent agreement with the experiments. The current modes
determined by these simulations show that the low frequency
resonances correspond to $\lambda$/2 current modes while the high
frequency resonances have 3$\lambda$/2 current modes. Retrieval of
the material's effective refractive index n, permittivity
$\varepsilon$ and permeability $\mu$ (see Fig. \ref{fig1}b) shows
that the structure has magnetic resonances at 4.5 and 13.5~GHz and
electric resonances at 6 and 16~GHz. The 6~GHz electric resonance,
which corresponds to excitation of a symmetric current mode in pairs
of rosettes (see Fig. \ref{fig1}b inset), is much stronger for LCP
($\text{Re}(\varepsilon)\approx-143$) than for RCP
($\text{Re}(\varepsilon)\approx-53$). Importantly the 4.5~GHz
magnetic resonance leads to a negative permeability for RCP of
$\text{Re}(\mu)\approx-1.9$, while the permeability for LCP shows
only weakly resonant behavior and remains positive. The negative
magnetic response for RCP arises from the excitation of
anti-symmetric currents in pairs of rosettes (see Fig. \ref{fig1}b,
inset). The RCP magnetic resonance is so strong that it leads to a
negative refractive index of $\text{Re}(n)\approx-0.9$ at about
4.5~GHz. Simultaneously the strong LCP electric resonance near 6~GHz
causes the refractive index to become negative.

Fig. \ref{fig2}b shows experimental results for the 4-layered form
of the metamaterial. The most striking difference compared to the
bi-layered case is a splitting of all resonances into LCP and RCP
resonances which leads to exceptionally high circular dichroism of
up to 20~dB with simultaneously low losses for one circular
polarization. This makes the metamaterial, which is only 1/5 of the
wavelength in thickness (around 13~GHz), well-suited as circular
polarizer. Furthermore the 4-layered material shows giant azimuth
rotation. True optical activity with a rotation angle of at least
$65^\circ$ is achieved at moderate losses and without making the
polarization state elliptical. The 6-layered version of the
metamaterial shows similarly strong polarization rotation and
circular dichroism.

All measurements were performed in an anechoic chamber using
broadband horn antennas (Schwarzbeck BBHA 9120D) and a vector
network analyzer (Agilent E8364B). Our numerical simulations were
done with CST Microwave Studio (Computer Simulation Technology GmbH,
Darmstadt, Germany) and Comsol Multiphysics, which both use a
frequency domain finite element method. Using the retrieval
procedure \cite{PRE_Smith_2005_EffPara}, we calculated the complex
effective parameters n, $\varepsilon$ and $\mu$ from the simulated
transmission and reflection.

In summary we demonstrate exceptionally strong gyrotropy in
metamaterial based on multi-layered meta-molecules consisting of
mutually twisted planar metal structures. Especially in forms of the
material composed of four or more layers we observe giant circular
dichroism and optical activity at moderate losses, which makes such
structures promising as sub-wavelength polarization rotators and
circular polarizers. Numerical simulations of the bi-layered
structure are in excellent agreement with the experimental results
and effective parameters derived from them show bands of negative
permeability for one and both circular polarizations and bands of
negative refractive index for each circular polarization.

\end{document}